\documentclass{fic-l}
\usepackage{epsfig}

\numberwithin{equation}{section}

\begin{document}

\title{Introduction to Multicanonical Monte Carlo Simulations}

%    Information for first author
\author{Bernd A. Berg}
%    Address of record for the research reported here
\address{Department of Physics\\ Florida State University\\ 
Tallahassee, Florida 32306, USA}
\thanks{Lecture given at the Fields Institute {\it Workshop on
Monte Carlo Methods}, October 1998. The author was supported 
in part by DOE Grants DE-FG05-87ER40319 and DE-FG05-85ER2500.}
%    Information for second author
% \author{Author Two}
% \address{Mathematical Research Section\\ School of Mathematical
% Sciences\\ Australian National University\\ Canberra ACT 2601,
% Australia}

%    General info
% \subjclass{Primary 54C40, 14E20; Secondary 46E25, 20C20}
% \date{July 2, 1991}

\dedicatory{E-mail: berg@hep.fsu.edu;\ 
Web: http://www.hep.fsu.edu/\~\,berg.}

\begin{abstract}

Monte Carlo simulation with {\it a-priori} unknown weights have
attracted recent attention and progress has been
made in understanding (i) the technical feasibility of such 
simulations and (ii) classes of systems for which such simulations
lead to major improvements over conventional Monte Carlo simulations.
After briefly sketching the history of multicanonical  calculations
and their range of application, a general introduction
in the context of the statistical physics of the $d$-dimensional 
generalized Potts models is given.  Multicanonical simulations yield 
canonical expectation values for a range of temperatures or any other 
parameter(s) for which appropriate weights can be constructed. We shall 
address in some details the question how the multicanonical weights 
are actually obtained. Subsequently miscellaneous topics related to
the considered algorithms are reviewed.  Then multicanonical studies 
of first order phase transitions are discussed and finally 
applications to complex systems such as spin glasses 
and proteins.

\end{abstract}

\maketitle

\section{Introduction and Summary}
 
One of the questions which ought to be addressed before performing a
large scale computer simulation is ``What are suitable weight factors 
for the problem at hand?'' Since the 1970s it has been expert wisdom 
that Monte Carlo (MC) simulations with a-priori unknown weight
factors are feasible and deserve to be considered
\cite{ToVa77}. With focus on narrow classes of applications, this
idea was occasionally re-discovered, for instance \cite{Ba87}. With
the work of ref.\cite{BeNe91,BeNe92} it became a more widely
accepted idea, which is employed for an increasing number of
applications in Physics, Chemistry, Structural Biology and
other areas~\cite{BeMe99}.

In 1988 the Italian APE\footnote{Array processor with emulator,
the collaboration named itself after their first dedicated computer.}
collaboration~\cite{Ba88} raised the question whether the $SU(3)$ 
deconfining phase transition in lattice gauge theory is really of
first order, as everyone believed, or possible of second order. The
first order nature of the transition was ultimately confirmed and
in our context it is only of interest that it was this question
which triggered the theoretical high energy physics community to
perform a number of numerical precision studies of first order phase
transitions. In particular difficulties to calculate the interfacial
tension were noted, as is reflected in the papers by Potvin and
Rebbi~\cite{PoRe89}, as well as Kajantie et al.~\cite{KaKa89},
where new methods were introduced and tested for 
the\footnote{Here and in the following $d$ denotes the dimension of 
the considered system.} $2d$ $7$-state Potts model
(and later applied to $SU(3)$ lattice gauge theory).
The author of this review was
involved in a number of studies of the $SU(3)$ deconfining
transitions, which cumulated in ref.\cite{AlBe92} with systematic
applications and developments of re-weighting techniques~\cite{FeSw88}
for $SU(3)$ gauge theory. In this context the papers with
Neuhaus~\cite{BeNe91,BeNe92} originated where we introduced the
{\it multicanonical} method and succeeded for the $2d$ 10-state Potts 
model to get accurate interface tension 
estimates out of Binder's~\cite{Bi82}
histogram technique. Looking back, it is to some extent astonishing
that we did not realize immediately the connection with Torrie and
Valeau's~\cite{ToVa77} {\it umbrella sampling} techniques. However,
Binder's histogram method (known to us by personal contacts) was by
then ten years old and more or less dormant due to the problem
of supercritical slowing down discussed below. It was just not a
natural idea to search the literature under the assumption that
everyone may have overlooked an existing numerical method that gets
it working. The umbrella method had stayed confined to small
expert circles. The major reason for this was, most likely, that
non-experts could never figure out how to get the {\it a-priori}
unknown weight factors in the first place. With respect to this 
point the multicanonical papers initiated progress beyond
simply making the method popular.

We are here concerned with MC simulations of the Gibbs canonical 
ensemble, see the next section for preliminaries. 
Multicanonical simulations calculate canonical expectation values 
in a temperature range, whereas conventional canonical 
simulations~\cite{Me53} calculate at a fixed temperature $T$ and
can, by re-weighting techniques, only be extrapolated to a
vicinity of this temperature. Interest into re-weighting
techniques, an idea which can be traced back to the late 
1950's~\cite{SaJa59}, exploded with the paper by Ferrenberg
and Swendsen~\cite{FeSw88}. Apparently, the time was then 
right for the radical step of transgressing entirely beyond 
conventional, canonical MC simulations. Still, a lucky accident 
helped the rapid acceptance of the multicanonical method. Namely,
for the $2d$ 7-state Potts model ref.~\cite{PoRe89,KaKa89} had
produced interface tension estimates which were by an entire order
of magnitude larger than the multicanonical estimates~\cite{JaBe92}
for the same model. Normally, such numerical discrepancies are
difficult to resolve. However, for the $2d$ $q$-state Potts model a
miracle happened: Building on results of~\cite{KlSc89,Kl90,BuWa93}
the {\it exact} $2d$ Potts model interface tensions were
derived~\cite{BoJa92} shortly after the simulations were completed.
The exact results and the multicanonical
estimates were found to be in excellent agreement and the previous
controversy converted to a significant boost for the new method.

In the application to calculations of interface tensions the 
emphasis is on enhancing rare configurations. Canonical MC
simulations~\cite{Me53} sample configurations $k$ with the
Boltzmann\footnote{Recently Tsallis~\cite{Ts88} weights 
received also attention~\cite{HaOk99b}.} weights
\begin{equation} \label{wcanonical}
 \widehat{w}_B (k) = {w}_B (E^{(k)}) = e^{-\beta E^{(k)} }
\end{equation}
where $E^{(k)}$ is the energy of configuration $k$, $\beta = 1/T$ and
units are chosen such that $k_B=1$ holds for the Boltzmann constant.
The resulting probability distribution for the energy is
\begin{equation} \label{Pcanonical}
 P(E) = c_{\beta}\, w_B (E) = c_{\beta}\, n(E)\, e^{-\beta E}
\end{equation}
where $n(E)$ is the spectral density, \textit{i.e.} the number of 
configurations of energy $E$. The normalization constant $c_{\beta}$
is needed to ensure $\sum_E P(E) = 1$.  Let $L$ characterize
the lattice size (for instance $N=L^d$ spins). For first order
phase transitions pseudocritical points $\beta^c (L)$ exist such
that the energy distributions $P(E)=P(E;L)$ become double peaked and 
$\beta^c(L)$ can be chosen such that the maxima at 
$E^1_{\max} < E^2_{\max}$ are of equal height 
$P_{\max} = P (E^1_{\max}) = P (E^2_{\max})$. In-between these 
values a minimum is located at some energy $E_{\min}$.
Configurations at $E_{\min}$ are exponentially suppressed like
\begin{equation} \label{Pmin}
P_{\min} = P (E_{\min}) = c_f\, L^p\, \exp ( - f^s A )
\end{equation}
where $f^s$ is the interface tension and $A$ is the minimal area 
between the phases, $A=2L^{d-1}$ for an $L^d$ lattice, $c_f$ and
$p$ are constants, $p=d-1$ in the capillary-wave
approximation~\cite{BrZi85,GeFi90,Mo91}. To determine the interface
tension one has to calculate the quantities
\begin{equation} \label{Itension}
f^s (L) = - {1\over A(L)}\, \ln R(L)\ ~~~~{\rm with}~~~~\ 
R(L) = {P_{\min} (L) \over P_{\max} (L) }
\end{equation}
and make a finite size scaling (FSS) extrapolation of $f^s(L)$ for
$L\to\infty$. However, for large systems a canonical MC
simulation~(\ref{Pcanonical}) will practically never visit 
configurations at energy $E=E_{\min}$ and estimates of the ratio 
$R(L)$ will be very inaccurate. The terminology \textit{supercritical
slowing down} was coined to characterize such an exponential
deterioration of simulation results with lattice size.
Ref.~\cite{BeNe92} overcame this problem by sampling, in an
appropriate energy range, with an \textit{approximation}
\begin{equation} \label{wmuca}
 \widehat{w}_{mu}(k)=w_{mu}(E^{(k)}) 
       = e^{-b(E^{(k)})\, E^{(k)} + a (E^{(k)})}
\end{equation}
to the weights
\begin{equation} \label{wspectral}
 \widehat{w}_{1/n} (k) = w_{1/n} (E^{(k)}) = {1\over n(E^{(k)})}\ .
\end{equation}
Here the function $b(E)$ is the \textit{microcanonical temperature} at 
energy $E$ and $a(E)$ is some kind of fugacity. Whereas approximations
to the weight factors\footnote{The first detailed discussion of the
connection of the multicanonical weights~(\ref{wmuca}) with the
inverse spectral density was given in~\cite{Be92}.}
$1/n(E)$ were already used in umbrella sampling \cite{ToVa77}, the
parameterization (\ref{wmuca}) in terms of the microcanonical
temperature $b(E)$ was introduced in~\cite{BeNe91} and is typical for
the multicanonical approach. The function $b (E)$ has a relatively
smooth dependence on its arguments which makes it a very useful
quantity when dealing with the weight factors.

With an approximation $w_{mu}(E^{(k)})$  to the weights
(\ref{wspectral}) one samples instead of the canonical energy
distribution $P(E)$ a new multicanonical distribution
\begin{equation} \label{Pmuca}
 P_{mu} (E) = c_{mu}\, n(E)\, w_{mu} (E) \approx c_{mu}\ .
\end{equation}
The desired canonical distribution (\ref{Pcanonical}) is obtained by
re-weighting
\begin{equation} \label{reweight}
 P (E) = {c_{\beta}\over c_{mu}}\, {P_{mu}(E) \over w_{mu} (E)}\,
         e^{-\beta E} .
\end{equation}
This relation is rigorous, because the weights $w_{mu}(E)$ used in
the actual simulation are exactly known. With the approximate relation
(\ref{Pmuca}) the average number of configurations sampled does not
longer depend strongly  on the energy and accurate estimates of the
ratio (\ref{Itension}) $R(L)=P_{\min}/P_{\max}$ become possible.
Namely for $i=1,2$ the equation
\begin{equation} \label{ratio}
 R(L) = R_{mu}(L)\,{w_{mu} (E^i_{\max})\,
 \exp (-\beta^c\,E_{\min}) \over w_{mu} (E_{\min})\,
 \exp (-\beta^c\,E_{\max})}\ ~~~~{\rm with}\ ~~~~
 R_{mu} (L) = {P_{mu} (E_{\min})\over P_{mu} (E^i_{\max})}
\end{equation}
holds and the statistical errors are those of the ratio $R_{mu}$ times
the exactly known factor. The errors of $R_{mu} (L)$ do not suffer
from supercritical slowing down, because in the multicanonical
simulation $E_{\min}$ is about as frequently visited as $E^1_{\max}$
or $E^2_{\max}$.

The attentive non-expert will raise the objection that the argument
appears to be circular. The weights (\ref{wspectral}) are
\textit{a-priori} unknown and would they be known we could choose
$w_{mu}(E^{(k)})=w_{1/n}(E^{(k)})$ and have
$ R_{mu}=P_{mu} (E_{\min})/P_{mu}(E_{\max}) = 1 $, {\it i.e.} there
is no need anymore to perform a simulation as $R(L)$ of equation
(\ref{ratio}) is then exactly known. The answer to this is that
the multicanonical method consists of two steps

\begin{enumerate}

\item Obtain a \textit{working estimate} $\widehat{w}_{mu}(k)$ of 
      the weights $\widehat{w}_{1/n}(k)$. Working estimate means
      that the approximation to
      (\ref{wspectral}) has to be good enough to ensure movement
      in the desired energy range, but deviations of $P_{mu} (E)$
      from the constant behavior (\ref{Pmuca}) by a factor of,
      say, ten are tolerable.

\item Perform a Markov chain MC simulation with the weights
      $\widehat{w}_{mu}(k)$. Canonical expectation values are found 
      by re-weighting to the Gibbs ensemble and standard jackknife
      methods \cite{Ef82,Be92a} allow reliable error estimates.

\end{enumerate}

For the $2d$ 10-state Potts model figure~1 reproduces thus obtained
multicanonical and re-weighted canonical energy histograms of 
ref.\cite{BeNe92}. The interface
tension estimate of ref.\cite{BeNe92} is depicted in figure~2.

\begin{figure}[tb]
\vspace{5pc}
 \centerline{\hbox{ \psfig{figure=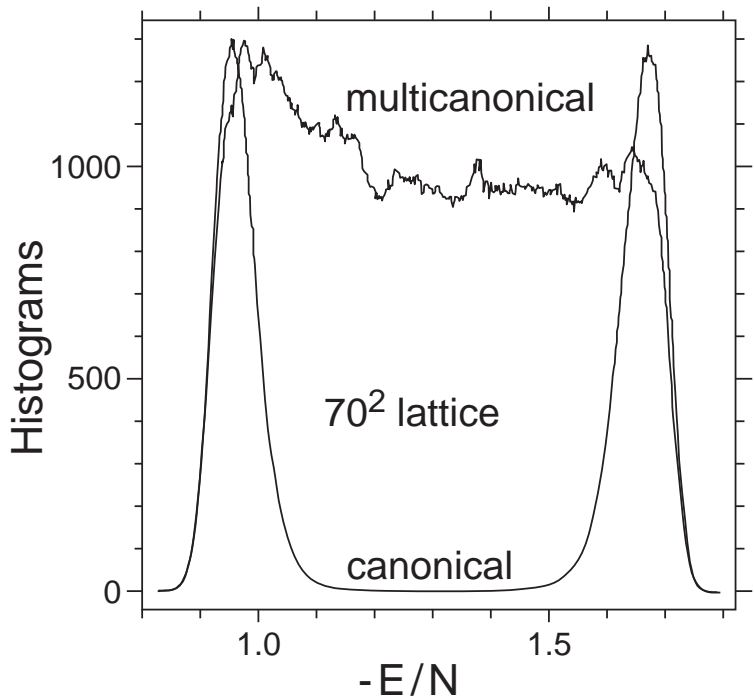,width=10cm} }}
 \caption{Multicanonical $P_{mu}(E)$ versus canonical $P(E)$ energy
 distribution as obtained in ref.\cite{BeNe92} for the $2d$ 10-state
 Potts model on a $70\times 70$ lattice.} \label{PE}
\end{figure}

\begin{figure}[tb]
\vspace{3pc}
 \centerline{\hbox{ \psfig{figure=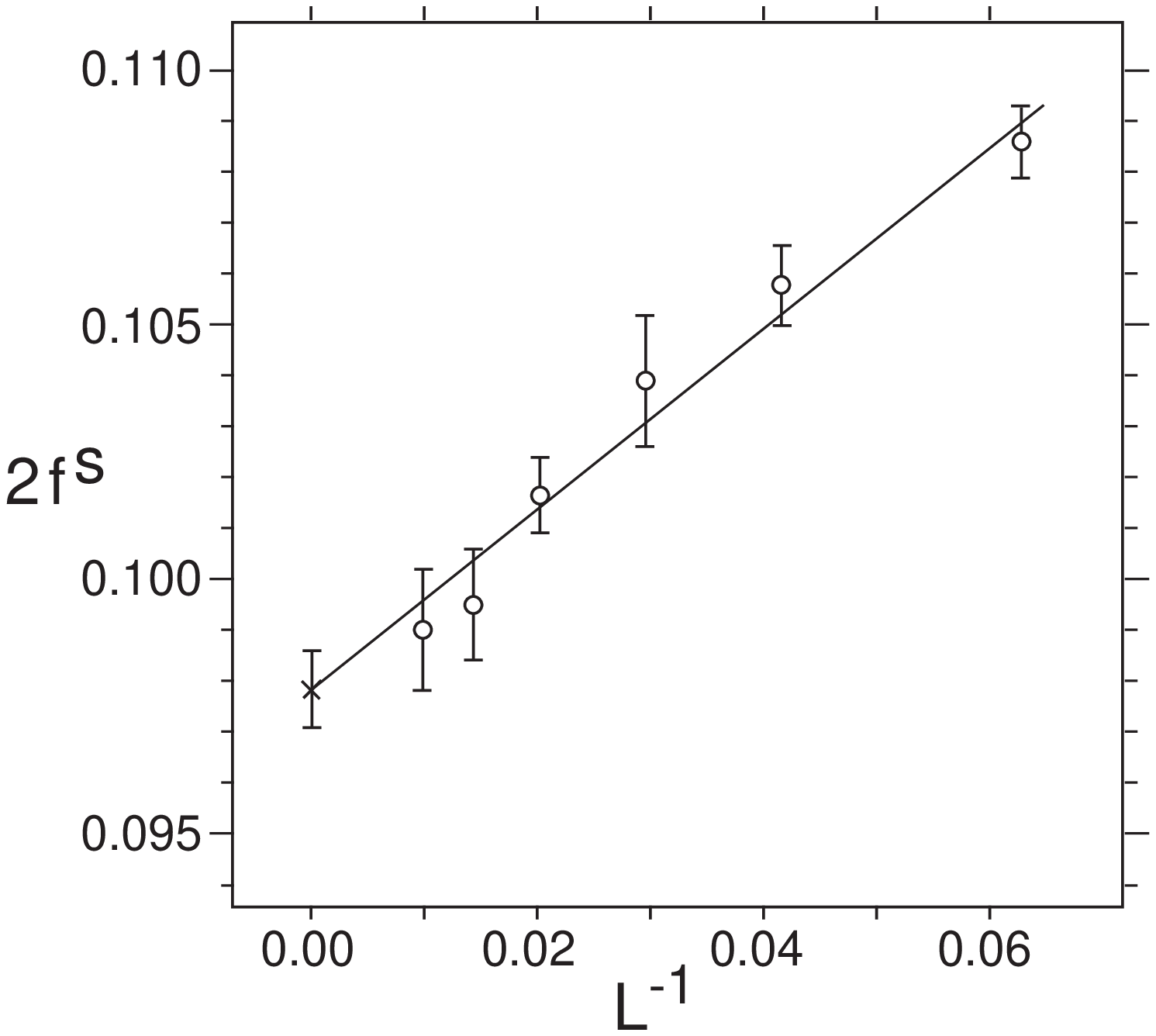,width=10cm} }}
 \caption{Interface tensions (\ref{Itension}) and their $L\to\infty$
 extrapolation of ref.\cite{BeNe92}.} \label{fs}
\end{figure}

Assume a researcher, familiar with the canonical MC method of
Metropolis et al.~\cite{Me53}, likes to get started with multicanonical
simulations. For canonical simulation the weights (\ref{wcanonical})
are exactly known. The first stumbling block is now to get the weight
factors $\widehat{w}_{mu}(k)$. For small to medium sized discrete
systems my recommendation
is to implement the general purpose recursion~\cite{Be98} which is
discussed in section~3. For first order phase transitions an
alternative is to rely on the FSS scaling behavior of the energy
distribution, see section~5. For large enough systems this approach
has advantages, but its applicability is limited to systems for which
a FSS theory of $P(E)$ exist.
Some people recommend constrained microcanonical MC simulations, which
are also shortly sketched at the end of section~5.

Once the weights $\widehat{w}_{mu}(k)$ are fixed the simulation is 
done by Markov chain MC in the usual way. The second complication,
which is minor compared to the first, is that canonical expectation
values of physical observables are no longer simple arithmetic 
averages of the generated (``measured'') values, but more involved 
equations have to be 
employed. This expands the programming efforts, but causes no real
difficulties. There is a major award for this additional work:
Multicanonical simulations allow for
immediate estimates of the {\it free energy} and the {\it entropy}, 
whereas their reconstruction from results of canonical simulations 
requires tedious integration techniques. These issues are reviewed
in the next section.

General properties of the method are discussed in section~4. This 
includes a discussion of its slowing down, {\it static} and 
{\it dynamic} aspects of the algorithm, and a summary  
of papers which have reported variants and new
progress.  In addition two related methods are sketched: 
\textit{parallel tempering} and \textit{random walk algorithms}. 
We then turn to applications in the next sections.

For first order phase transitions the multicanonical method is now 
well-estab\-lished and a number of applications are summarized in
section~5. Next, it has been realized early~\cite{BeCe92} that 
multicanonical simulations do not only allow to enhance the weights 
of relevant rare configurations, but can also be employed to improve 
the dynamics of the Markov process, {\it i.e.} its movement through 
configuration space in the presence of free energy barriers. 
Equation (\ref{Pmin}) is a particularly simple example of an explicitly
controlled free energy barrier. Of major interest in nowadays research
are complex systems, characterized by a rough free energy landscape for 
which an explicit parameterization is not known. The most important
complex systems are presumably proteins for which multicanonical
investigations have been pioneered by Hansmann and
Okamoto~\cite{HaOk93}. Applications of the multicanonical method to
these systems are summarized in section~6.
The lack of explicit parameterizations of the
free energy barriers makes complex systems difficult to study. For
simulations one has in principle two strategies: 
\begin{enumerate}
\item Enhance the suppressed configurations and move over the barriers.
\item Move around the barriers.
\end{enumerate}
Only the first strategy allows for an explicit calculation of barrier 
heights, because in the second configurations with barriers are still 
not sampled.  It is with respect to the second strategy, but not 
the first, that the multicanonical approach is in 
competition with the method of multiple Markov chains~\cite{Ge91}
(see {\it parallel tempering} in section~4).
The canonical weights are not changed in the latter method, 
configurations which are rare in the Gibbs ensemble (for instance 
configurations with interfaces) stay rare and the strength of the two
methods can be quite distinct, although a considerable overlap in
the range of potential applications exists.

Finally, conclusions and outlook are given in section~7. 

\section{Preliminaries}

MC simulations of systems described by the Gibbs canonical ensemble
aim at calculating estimators of physical observables $\mathcal{O}$.
For discrete systems (on a computer all systems are discrete) the
expectations values at temperature $\beta = 1/T$ are defined by
\begin{equation} \label{O}
 \mathcal{O} = \mathcal{O} (\beta) = 
 <\mathcal{O}^{(k)}> = Z^{-1} \sum_{k=1}^K
 \mathcal{O}^{(k)}\,e^{-\beta\,E^{(k)} }
\end{equation}
where
\begin{equation} \label{Z}
 Z = Z(\beta) = \sum_{k=1}^K e^{-\beta\,E^{(k)} }
\end{equation}
is the \textit{partition function}. The sum $k=1,\dots ,K$ goes over
all configurations (or microstates) of the system and $E^{(k)}$ is
the energy of microstate $k$.

Here we focus our discussion on generalized Potts models in an 
external magnetic field on $d$-dimensional hypercubic lattices. 
Without being overly complicated, these models are general enough to 
illustrate the essential features we are interested in.
In addition, various subcases of these models are by themselves of 
considerable physical interest. Conceptual generalizations to other 
models are straightforward, but technical complications arise in 
some cases. 

For generalized Potts models the energy of the system is given by
\begin{equation} \label{E}
 E^{(k)} = E_0^{(k)} + M^{(k)} 
\end{equation}
where
\begin{equation} \label{E0}
 E_0^{(k)} = - \sum_{<ij>} J_{ij}(q_i^{(k)},q_j^{(k)})\
 \delta (q_i^{(k)},q_j^{(k)})\ ~~~~{\rm with}~~\ \delta (q_i,q_j) 
 = \left\{ \begin{array}{c} 1\ {\rm for}\ q_i=q_j \\
         0\ {\rm for}\ q_i\ne q_j \end{array}  \right.
\end{equation}
and
\begin{equation} \label{M}
 M^{(k)} = H \sum_{i=1}^N \delta (1,q_i^{(k)})\ .  
\end{equation}
In equation (\ref{E0}) the sum $<ij>$ is over the nearest neighbour
lattice sites and $q_i^{(k)}$ is the state of configuration $k$ at
site $i$. For the $q$-state Potts model $q^{(k)}_i$ takes on the
values $1,\dots ,q$. The $J_{ij} (q_i,q_j)$, $(q_i=1,\dots ,q;\,
q_j=1,\dots, q)$ functions denote exchange coupling constants between
the states at site $i$ and site~$j$. For
\begin{equation} \label{Jij}
 J_{ij} (q_i,q_j) \equiv J > 0~~\ ({\rm conventionally}\ J=1) 
\end{equation}
the original Potts model is recovered and $q=2$ becomes
equivalent to the Ising ferromagnet. Potts glasses~\cite{Bi97} and
Ising spin glasses~\cite{EdAn75} are obtained when the exchange
constants are quenched random variables. Other choices of the
$J_{ij}$ include anti-ferromagnets and the fully frustrated Ising
model~\cite{Vi77}.

The sum in equation (\ref{M}) goes over all $N$ sites $i$ of the
lattice. The external magnetic field is chosen to interact with
the state $q_i=1$ at each site $i$, but not with the other states. 
Each configuration (microstate of the system) $k$ defines a particular
arrangements of all states at the sites and, vice versa, each
arrangement of the states at the sites determines uniquely a
configuration:
\begin{equation} \label{k}
 k = \{ q_1^{(k)}, \dots , q_N^{(k)} \} \ . 
\end{equation}
As there are $q$ possible states at each site, the total number of
microstates is
\begin{equation} \label{K}
 Z(0) = K = q^N\ ,  
\end{equation}
where we have used the definition (\ref{Z}) of $Z$. Already for
moderately sized systems $q^N$ is an enormously large number. If
$L_n$, $(n=1,\dots ,d)$ are the edge lengths of the lattice, the
number of sites is given by
\begin{equation} \label{N}
 N = \prod_{n=1}^d L_n\ ,  
\end{equation}
\textit{i.e.} $N=L^d$ for a symmetric lattice with $L_1=\dots =L_N=L$.

Metropolis et al.~\cite{Me53} introduced {\it importance sampling} for
the Gibbs canonical ensemble\footnote{For an introduction to the
canonical ensemble and statistical physics in general see for
instance~\cite{Hu87}.}. Their original approach imitates the thermal
fluctuations of nature and weights configurations with the Boltzmann
factors (\ref{wcanonical}). The implementation relies on a Markov
process and generalizes immediately to arbitrary weights
$\widehat{w}(k)$. Given a configuration $k$, new configurations
$k'$ are proposed with probabilities of an explicitly known
transition matrix
\begin{equation} \label{p0}
 p^0(k',k) = p^0(k,k'),\ \sum_{k'} p^0(k',k)=1\ .
\end{equation}
The symmetry condition is known as {\it detailed balance} and
can be replaced by the somewhat weaker condition of balance,
see for instance~\cite{Bi76}. The newly proposed configuration
$k'$ is accepted with the probability
\begin{equation} \label{paccept}
 p^a(k',k) = \min [1, \widehat{w}(k',k)]\ ~~~~{\rm with}~~~~\
 \widehat{w} (k',k) = {\widehat{w} (k') \over \widehat{w} (k)}
\end{equation}
and otherwise rejected.
Here $\widehat{w}(k)$ can be the Boltzmann factors~(\ref{wcanonical}), 
the multicanonical weights~(\ref{wmuca}) or any
other function of $k$. Putting equations (\ref{p0}) and (\ref{paccept})
together, the transition probability for $k\to k'$ becomes
\begin{equation} \label{p_transition}
 p (k',k) = p^0(k',k)\, p^a(k',k)\ {\rm for}\ k'\ne k\
 ~~~~{\rm and}~~~~\ p(k,k) = 1 - \sum_{k'\ne k} p(k',k)\ .
\end{equation}
Note that rejected proposals $k\to k'$ have to be counted as $k\to k$
transitions. It is a well-known beginner's mistake to undercount the
at hand configuration $k$.

It was first proven in~\cite{Me53} that the procedure
(\ref{p_transition}) generates configurations $k$ with the
desired weights $w(k)$ provided the property of {\it ergodicity}
holds: Every configurations has to be reachable in a finite number
of steps.

For generalized Potts models implementation of the Metropolis
algorithm is straightforward. The transition matrix $p^0(k',k)$
can be defined by the following procedure
\begin{enumerate}
\item Pick one site $i$ at random, {\it i.e.} with probability $1/N$.
\item Assign one of the states $1,\dots,q$ to $q_i'$, each with
      probability $1/q$.
\end{enumerate}
These rules account for
\begin{equation} \label{Potts_p0}
 p^0(k',k) =  \left\lbrace \begin{array}{c}
 1/q\ {\rm for}\ k'=k,\ \ \ \ \ \ \ \ \ \ \ \ \ \ \\
 (q-1)/(qN)\ {\rm for\ each\ configuration}\ k'\
 {\rm with}\ q_i'\ne q_i\ {\rm at\ one\ site}\ i.
 \end{array} \right. 
\end{equation}
The new configuration $k'$ is then accepted or rejected according to 
the rule (\ref{paccept}). Ergodicity is fulfilled, because with
(\ref{Potts_p0}) every configuration can be reached
in a minimum number of $N$ update steps. The random choice of
a site insures detailed balance. A similar procedure where
states are updated in the systematic order of a permutation
$\pi_1,\dots,\pi_N$ leads still to the desired distribution.
Although detailed balance is then violated the weaker condition
of balance still holds. 

The Metropolis process needs to run some while to establish
{\it equilibrium} with respect to the {\it ensemble} defined by 
the weights with which it samples, see for instance~\cite{Bi76}
for an introduction and \textit{coupling from the 
past}~\cite{PrWi98} is a recent rigorous approach. In the 
following we assume that equilibrium has been reached. Subsequently
the Metropolis process generates a number of equilibrium configurations
$k_i$, $i=1,\dots,n$ on which observables $\mathcal{O}$ are 
{\it measured} similarly as in real experiments. Let us focus on the 
internal energy $E$. When Boltzmann weights 
$\widehat{w}_B(k)$ (\ref{wcanonical}) 
are used the Metropolis process generates $E^i=E^{(k_i)}$ values 
in the canonical ensemble and the estimator $\overline{E}$ of 
$\mathcal{O}=E$ in~(\ref{O}) is simply the arithmetic average
\begin{equation} \label{overlineEcanonical}
 \overline{E} = {1\over n} \sum_{i=1}^n E^i\ .
\end{equation}
When other weights are used this is no longer true. After equilibrium
is reached, a Metropolis process with the weights
$\widehat{w}_{mu}(k)$ (\ref{wmuca}) will generate configurations in 
the {\it multicanonical ensemble}. To obtain estimators of canonical
expectation values (\ref{O}) one has to re-weight to the canonical
ensemble and equation (\ref{overlineEcanonical}) becomes 
\begin{equation} \label{overlineEmuca}
 \overline{E} (\beta ) = 
 \sum_{i=1}^n E^i\, w^{-1}_{mu}(E^i)\, \exp (-\beta E^i)
 \left/ \sum_{i=1}^n w^{-1}_{mu}(E^i)\, \exp (-\beta E^i) \right. \ .
\end{equation}
It is instructive to compare the histograms
$H_i(E)=H_{\beta_i}(E)$, $(i=1,2)$ of two canonical simulations
($\beta_1 < \beta_2)$
with the histogram $H_{mu}(E)$ generated by a multicanonical
Metropolis process. These histograms are
estimators of the corresponding probability densities $P(E)$
(\ref{Pcanonical}) and $P_{mu}(E)$ (\ref{Pmuca}). A typical
situation is depicted in figure~3. Histograms from canonical
simulations at sufficiently distinct temperatures will not 
overlap, whereas a histogram from a multicanonical simulation
can bridge the entire range and allows to calculate canonical
expectation values for the range
$\beta_1 \le \beta \le \beta_2$ (and, in the case depicted, 
also for some $\beta > \beta_2$).

\begin{figure}[tb]
 \centerline{\hbox{ \psfig{figure=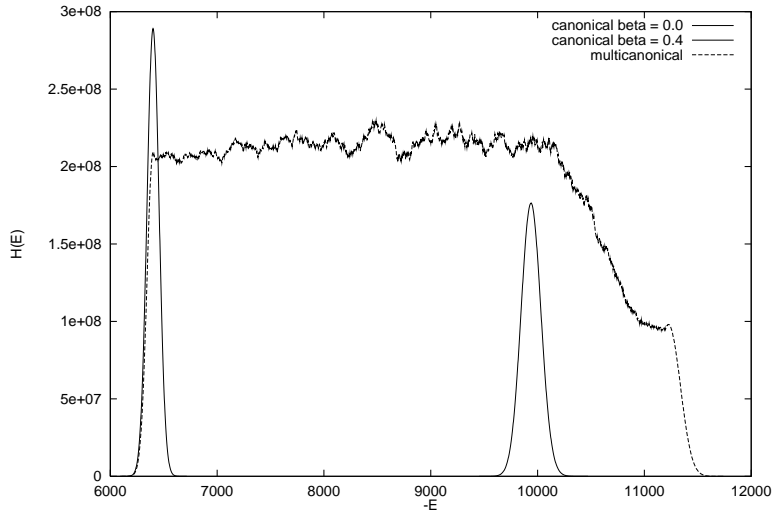,width=10cm} }}
 \caption{Simulation of a $2d$ Ising model on a $80^2$ lattice:
 Canonical histograms $H_1(E)$ and $H_2(E)$ from simulations at
 $\beta_1=0$ and $\beta_2=0.4$ versus a multicanonical histogram
 $H_{mu}(E)$ which includes their ranges.}
\label{histograms}
\end{figure}

Obviously the re-weighting (\ref{overlineEmuca}) can only work
in a temperature range where the multicanonical histogram includes
the canonical histogram from a simulation at that temperature. For
insufficient statistics (or a false choice of multicanonical
weights) the entries from outside the $H_{\beta}(E)$ range may lead
to spurious estimates of $\overline{E}(\beta)$ (\ref{overlineEmuca}).
This is an example of a {\it bias} problem. For dealing with such
problems the jackknife method, see for instance~\cite{Ef82,Be92a}, 
is recommended. Already for the statistical analysis of results 
from canonical simulations this should really be the method of first 
choice and as such be taught in introductory books on MC simulations. 
For multicanonical simulations bias problems tend to become more 
subtle and the use of jackknife estimators becomes often a must.

It is well-known~\cite{Bi76} that a canonical MC simulation at 
temperature $\beta = 1/T$ provides estimates for many but not all
thermodynamical quantities of interest. Obtained are estimators 
for the internal energy $E$, quantities related to its derivatives, 
like the specific heat
\begin{equation} \label{C_V}
 C_V = - {\partial E\over \partial T} = \beta^2\, {\partial E
 \over \partial \beta} = \beta^2\, ( <E^{(k)} E^{(k)}> - 
 <E^{(k)}> <E^{(k)}> )\ , 
\end{equation}
correlation functions, the magnetization and many others. However,
canonical simulations fails with respect to a number of important
quantities, most prominently the partition function $Z(\beta)$
itself and, related to it, for the Helmholtz free energy 
\begin{equation} \label{F}
 F (\beta ) = -\beta^{-1}\, \ln Z (\beta )
\end{equation}
and the entropy\footnote{This is the canonical entropy which has
to be distinguished from the microcanonical entropy used in the
next section.}
\begin{equation} \label{S}
 S = { E - F \over T}\ . 
\end{equation}
The reason is that the histograms $H(E)$ of canonical sampling yield
the shape of the probability density $P(E)$ but do not estimate the
normalization constant $c_{\beta}$ (\ref{Pcanonical}) which, due to
$c_{\beta} \sum_E n(E)\,\exp (-\beta E)=1$, is
\begin{equation} \label{cbeta}
 c_{\beta} = 1 / Z (\beta )\ .
\end{equation}
Instead the normalization constant for the generated histogram is
simply the inverse (arbitrary) number of generated configurations.
In contrast to this multicanonical simulations allow to calculate 
$Z(\beta )$ by exploiting that $Z(0)=K=q^N$ is known (\ref{K}). After
the multicanonical simulation the estimator for the canonical 
probability density (\ref{reweight}) is
\begin{equation} \label{Hreweight}
 P (E) = {c_{\beta}\over c^n_{mu}}\, {H_{mu}(E) \over w_{mu} (E)}\,
         e^{-\beta E} 
\end{equation}
where the superscript $^n$ on $c^n_{mu}$ indicates that this is the
constant needed when using the histogram which relies on $n$
configurations generated in the multicanonical ensemble. Assume that
the $\beta$-range for which re-weighting is valid includes $\beta=0$.
The constant $c^n_{mu}$ follows then from the known result
$c_0=1/Z(0)=1/K$. Once $c^n_{mu}$ is known $c_{\beta}$ (and hence
the partition function) follows from the normalization $\sum_E P(E)=1$,
where $\beta$ has of course to stay in the admissible re-weighting
range.

\section{Multicanonical Recursion}

In this section a recursion~\cite{Be98} is presented which allows 
to obtain working estimates $w_{mu}(k)$ of the weight
factors (\ref{wspectral}). Essentially, it is a simplified and
improved version of a recursion which was first published
in~\cite{Be96} after being extensively tested in
spin glass simulations~\cite{BeHa94}.

It is recommended to start 
the recursion in the disordered phase of a system, where  the system 
moves freely under MC updates. The {\it zeroth} simulation is done
with the weights
\begin{equation} \label{w0}
w^0  (k) = 1 \ {\rm for\ all}\ k\, .
\end{equation}
The most obvious recursion towards the weights (\ref{wspectral}) goes
as follows: Simulation $n$, $(n=0,1,2,...)$ is carried out with the 
estimate $w^n (k) $  and yields the histogram $H^n (E)$. Estimate
$n+1$ for the weight factors is then given by

\begin{equation} \label{naive}
w^{n+1} (k) = { w^n (k) \over H^n (E^{(k)}) }\, . 
\end{equation}
This simple-minded approach fails due to a number of  difficulties

\begin{enumerate}
\item What to do with histogram entries $H^n (E) = 0$ or small?
\item Each $H^n (E)$ calculation starts with zero statistics. Assume,
someone has given us the exact weights (\ref{wspectral}) and we use 
$w^0 (k)=w_{1/n}(k)$: The next estimate $w^1 (k)$ will be worse.  A 
noisy left-over of $w_{1/n} (k)$.
\item The initial weights $w^0(k)=1$ correspond to temperature 
infinity and are bad in the limit $E^{(k)}\to E_g$, where $E_g$ is the
groundstate of the systems which is approached in the zero temperature 
limit.
\end{enumerate}

The recursion derived in the following overcomes these difficulties.
Let us first discuss the relationship \cite{BeNe91,Be96} of the
weights (\ref{wmuca}) with the microcanonical temperature $b (E)$ and
the fugacity $a(E)$, because it turns out to be advantageous to
formulate the recursion in terms of these quantities. We have
\begin{equation}
w (k) = e^{-S(E^{(k)})} = e^{-b(E^{(k)})\, E^{(k)} + a(E^{(k)})}\, 
\end{equation}
where $S(E)$ is the microcanonical entropy and,  by definition,
\begin{equation}
 b (E) = {\partial S (E) \over \partial E}\ .
\end{equation}
This determines the fugacity function $a (E)$ up to an 
(irrelevant) additive constant: We consider the case
of a discrete minimal energy $\epsilon$ and choose 
\begin{equation} \label{bE}
 b(E) = \left[ S(E+\epsilon) - S(E) \right] / \epsilon\ .
\end{equation}
The identity $S(E) = b(E)\, E - a(E)$ implies
$$ S(E) - S(E-\epsilon) = b(E) E - b(E-\epsilon) (E-\epsilon ) -
a(E) + a(E-\epsilon)\ .$$
Inserting $\epsilon\, b(E-\epsilon) = S(E) - S(E-\epsilon)$ yields 
\begin{equation}
 a(E-\epsilon) =  a(E) + \left[ b(E-\epsilon)-b(E) \right]\, E
\end{equation}
and $a(E)$ is fixed by defining $a(E_{\max})=0$.
In summary, once $b(E)$ is given, $a(E)$ follows for free.
The starting condition (\ref{w0}) becomes (other $b^0 (E)$ choices 
are of course possible)
\begin{equation}
 b^0 (E)= a^0 (E) = 0\, .
\end{equation}
To avoid $H(E)=0$ we replace for the moment
\begin{equation} \label{hatH}
H(E)\to {\hat H(E)} = \max\, [h_0,H(E)]\, ,
\end{equation}
where $0 < h_0 < 1$. Our final equations will allow for the limit
$h_0 \to 0$. With this replacement we translate equation~(\ref{naive})
into an equation for $b(E)$. Subscripts $_0$ are used to indicate
that those quantities are not yet our final estimators from the
$n^{th}$ simulation. Let
$$w^{n+1}_0(E)=e^{-S^{n+1}_0(E)}=c\, {w^n(E) \over \hat{H}^n(E)}\, ,$$
where the (otherwise irrelevant) constant $c$ is introduced to ensure 
that $S^{n+1}_0(E)$ can be an estimator of the microcanonical entropy. 
It follows
\begin{equation} 
 S^{n+1}_0(E) = - \ln c+S^n(E)+\ln \hat{H}^n(E)\ .
\end{equation}
Inserting this relation into (\ref{bE}) gives
\begin{equation} \label{b0} 
 b^{n+1}_0 (E) = b^n (E) + [ \ln \hat{H}^n(E+\epsilon) -
 \ln \hat{H}^n(E) ] / \epsilon
\end{equation}
The estimator of the variance of $b^{n+1}_0(E)$ is obtained from 
$$\sigma^2 [ b^{n+1}_0(E)] = \sigma^2[ b^n (E) ] +
  \sigma^2 [ \ln \hat{H}^n(E+\epsilon)] / \epsilon + 
  \sigma^2[  \ln \hat{H}^n(E) ] / \epsilon\ . $$
Now $\sigma^2[b^n (E)]=0$ as $b^n(E)$ is the fixed function used in
the $n^{th}$ simulation and the fluctuations are governed by
the sampled histogram $H^n=H^n(E)$
$$ \sigma^2[ \ln (\hat{H}^n )] = \sigma^2[ \ln (H^n)] =
\left[ \ln (H^n + \triangle H^n) - \ln (H^n) \right]^2 $$
where $\triangle H^n$ is the fluctuation of the histogram, which
is known to grow with the square root of the number of entries
$\triangle H^n \sim \sqrt{H^n}$. Hence,
\begin{equation} \label{sigma2_b} 
 \sigma^2[ b^{n+1}_0(E)]
={c'\over H^n(E+\epsilon)}+{c'\over H^n(E)}\ ,
\end{equation}
holds where $c'$ is an unknown constant and this equation emphasizes
that the variance is infinite when there is zero statistics,
{\it i.e.} $H^n(E)=0$ or $H^n(E+\epsilon)=0$. The statistical weight
for $b^{n+1}_0 (E)$ is inversely proportional to its variance and the
over-all constant is irrelevant. Choosing a convenient over-all
constant we get
\begin{equation}
g^n_0 (E) = {H^n (E+\epsilon)\ H^n (E) \over
H^n (E+\epsilon) + H^n (E)}\, .
\end{equation}
Note that $g^n_0(E)=0$ for $H^n(E+\epsilon)=0$ or $H^n(E)=0$.
The $n^{th}$ simulation was carried out using $b^n (E)$. It is
now straightforward to combine $b_0^{n+1} (E)$ and 
$b^n (E)$ according to their respective statistical weights
into the desired estimator:
\begin{equation} \label{bn}
b^{n+1} (E) = \hat{g}^n (E)\,  b^n (E) +
                \hat{g}^n_0 (E)\, b^{n+1}_0 (E)\, ,
\end{equation}
where the normalized weights
$$ \hat{g}^n_0 (E) = {g^n_0(E) \over g^n(E) + \hat{g}^n_0 (E)}\
{\rm and}\ \hat{g}^n (E) = 1 - \hat{g}^n_0 (E) $$
are determined by the recursion 

\begin{equation}
g^{n+1} (E) = g^n(E) + g^n_0(E),\ g^0(E)=0\, .
\end{equation}
We can eliminate $b^{n+1}_0 (E)$ from equation~(\ref{bn}) by
inserting its definition (\ref{b0}) and get
\begin{equation} \label{recursion}
 b^{n+1}(E) = b^n (E) + \hat{g}^n_0(E) \times 
 [ \ln \hat{H}^n(E+\epsilon)-\ln \hat{H}^n(E)] / \epsilon
\end{equation}
Notice that it is now save to perform the limit
$$ h_0\to 0 $$
in the definition of $\hat{H}$ (\ref{hatH}). Namely,
$\hat{g}^n_0(E)=0$ for $H^n(E)=0$ or $H^n(E+\epsilon)=0$ and
$$ \hat{g}^n_0 \times \lim_{h_0\to 0} 
 [ \ln \hat{H}^n(E+\epsilon)-\ln \hat{H}^n(E)] $$
is well-defined for all $H^n(E)$. 

In contrast to the naive recursion (\ref{naive}) the recursion
(\ref{recursion}) includes the entire information assembled.
Frequent iterations are allowed, implying increased stability and 
decreased CPU time consumption. In addition it is recommended to
implement a suitable $b (E)$ guess for the not yet covered
$E\to E_{\min}$ energy values. Figure~4 depicts how the recursion
works for the $2d$ 10-state Potts model (magnetic field $H=0$)
on an $80^2$ lattice. The straight horizontal lines indicate the
$b(E)$, $E\to E_g$, guess used after the corresponding recursion
step and the noise at their off-spring points comes from the
last recursions done before the snapshot was taken. Subsequent 
recursion steps were separated by one thousand MC sweeps.

% The results of this figure were obtained with the extrapolation
% \begin{equation} \label{bextra}
% b^n(E) = b_{\max}^n\ ~~{\rm for}~~\ E \le (3\,E^n_{min}
%        + E^n_{cut} ) / 4
% \end{equation}
% where $E^n_{min}$ is the smallest energy for which a $\hat{g}^n(E)$
% is non-zero (it has nothing to do with $E_{\min}$ of
% equation~(\ref{Pmin})) and
% \begin{equation} \label{bmax}
% b^n_{\max} = \sum_{E=E^n_{min}}^{E^n_{cut}} \hat{g}^n(E)\, b(E)
% \left/ \sum_{E=E^n_{min}}^{E^n_{cut}} \hat{g}^n(E) \right. \ .
% \end{equation}
% Here $E^n_{cut}$ is a parameter which allows to limit
% low statistics fluctuations. For the simulations of
% figure~4 it was set to be the smallest energy for which
% $$ p\le \sum_{E=E^n_{min}}^{E^n_{cut}} \hat{g}^n(E) \left/ 
%   \sum_{E=E^n_{min}}^0 \hat{g} \right. ~~{\rm with}~~ 
%   p = {5\over n+9} $$
% holds and subsequent recursion steps were separated by one
% thousand MC sweeps.

\begin{figure}[tb]
 \centerline{\hbox{ \psfig{figure=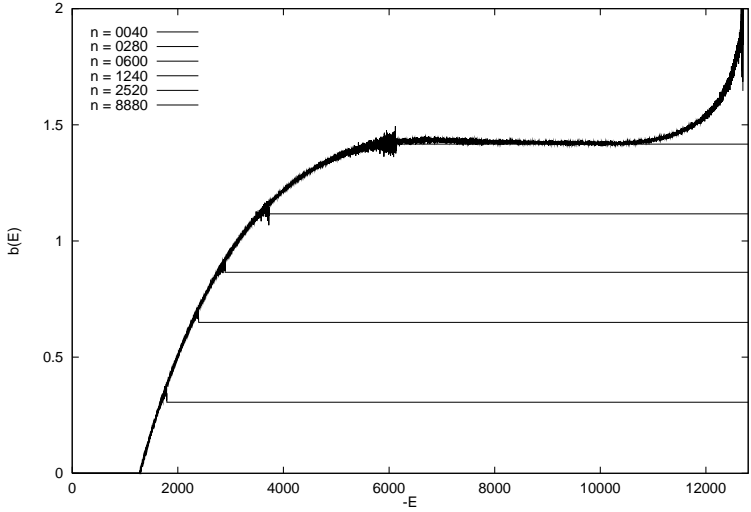,width=10cm} }}
 \caption{Recursion $b^n(E)$ for the 10-state Potts model on an
 $80^2$ lattice.}\label{protein}
\end{figure}

Finally, equation~(\ref{recursion}) can be converted into a direct 
recursion for ratios of the weight factor neighbors. We define
\begin{equation} 
R^n (E) = e^{\epsilon b^n (E)} = {w^n(E) \over w^n (E+\epsilon)}
\end{equation}
and get~\cite{BeJa98}
\begin{equation}
R^{n+1} (E) = R^n (E)\, \left[ {\hat{H}^n (E+\epsilon) \over
              \hat{H}^n (E) } \right]^{\hat{g}^n_0(E)}\, .
\end{equation}
This version is of interest when RAM limitations are an issue.

\section{Properties of the Algorithm and Miscellaneous Topics}

Before we come to the two major fields of applications,
first order phase transitions and complex systems, it
is helpful to summarize a number of algorithmic issues.
A reader who is not interest in more technical points may 
want to skip this section.

\subsection{Slowing down} 

Our typical situation is
$$ E_{\max} - E_{\min} \sim V $$
The optimum for a flat energy distribution is given by a 
random walk in the energy~\cite{BeNe91}. This implies a CPU time 
increase
$$ \sim V^2 $$
to keep the number of $E_{\max}\to E_{\min}\to E_{\max}$
transitions constant.  The recursion (\ref{recursion})
needs an additional $\sim V^{0.5}$ (optimum)
attempts to cover the entire range. It follows:
$$ {\rm slowing\ down}\ \sim\ V^{2.5}\ {\rm or\ worse.}$$
For first order phase transitions (see the next section)
a {\it recursion alternative} is patching of overlapping
constraint MC simulations:
$$ {\rm number\ of\ (fixed\ size)\ patches}\ \sim\ V\, .$$
When results can be obtained by keeping the number of updates per
spin (sweeps) in each patch constant, another CPU factor $\sim V$
follows. In this case we can get:
$$ {\rm optimal\ performance}\ \sim\ V^2\, .$$
However, in practice this approach may face ergodicity problems.

\subsection{Static and dynamic aspects of the algorithm}

The actual performance of the algorithm tends to be close
to the optimal performance, when one focuses on static
applications. Here {\it static} means that some canonically
rare configurations have to be enhanced. In {\it dynamical}
applications one uses the method to by-pass rare configurations
without actually enhancing then. Whether the slowing down is
still close to the optimum or not depends then on whether
a suitable parameterization of the problem can be found and 
for complex systems this is an open issue.

{\it Static Examples:}

\begin{itemize}

\item Magnetic field driven first-order phase transitions:
Configurations with zero (or small) magnetic fields are
exponentially suppressed at low temperatures and they
exhibit domain walls.

\item Temperature driven first-order phase transitions:
Configurations with domain walls are exponentially suppressed.

\item These rare configurations can be
      enhanced~\cite{BeNe92,BeHa93a}.

\end{itemize}

{\it Dynamic Examples:}

\begin{itemize}

\item The following transitions my be induced by multicanonical
simulations which include the high temperature region~\cite{BeCe92}:

\item Low temperature transitions between magnetic states
(for instance the up-down states of the
Ising model below the Curie temperature, ...).

\item Transition between low temperature states in systems
with conflicting constraints: Spin glasses, proteins, the traveling
salesman problem, ...

\end{itemize}

\subsection{Variants of the multicanonical methods} {\it Multicanonical}
refers to calculations of canonical expectation values for a temperature
range and re-weighting has to be done in the internal energy. Similarly,
other physical quantities can be considered, {\it e.g.} 
{\it multimagnetical} \cite{BeHa93a} refers to simulations
which give results for a certain range of the magnetic field.
A variant for cluster updates due to Janke and
Kappler~\cite{JaKa95} is called {\it multibondic}. Recently
the {\it multi-overlap} algorithm was introduced~\cite{BeJa98a}
which focuses on the Parisi order parameter for spin glasses. 
All these algorithms fall into the general class of {\it umbrella
sampling} methods, which knows similar distictions, like
temperature-scaling MC~\cite{ToVa77} and the later introduced
density-scaling MC~\cite{Va91}.

\subsection{Technical progress} A number of papers have reported 
further progress along the lines of multicanonical algorithms. For
asymmetric distributions Borgs and Kappler~\cite{BoKa92} suggest
to use an equal weight instead of the equal height criterium 
employed in equation~(\ref{Itension}). Hesselbo and 
Stinchcombe~\cite{HeSt95} made an attempt to optimize the 
weights and propose
\begin{equation} \label{wHS}
 w_{HS}(E) = 1 / \sum_{E' \le E} n(E) 
\end{equation}
instead of (\ref{wspectral}). 
Their arguments cover the static situation (pertinence), but as
one has no {\it a-priori} control over the dynamics of the
systems, the question of optimal weights remains one of trial
and error.  Combining 
multicanonical with multigrid methods has been explored by Janke 
and Sauer~\cite{JaSa94}. For molecular dynamics, Langevin and hybrid 
MC variants see Hansmann et al.~\cite{HaOkEi96} and, with emphasize 
on lattice gauge theory, Arnold et al.~\cite{ArLi98}. Connections
with adaption and linear response theory were explored 
in~\cite{MuOy96} and optimization
of estimators is discussed in~\cite{FiPiSi99}. Occasionally
attempts have been made to use bivariate multicanonical 
weighting~\cite{HiNa97}, with most spectacular results claimed
recently~\cite{Hat99}. Concerning the latter reference, it
appears to be too early for a final judgement. In general
numerical methods tend to become unstable when the parameter
space becomes too big.

\subsection{Parallel tempering}

The developments sketched above have to be distinguished from 
the the {\it parallel tempering} approach which is for some
applications the major competing method and should by no means
be confused with multicanonical sampling.

The first paper on parallel tempering is due to Geyer~\cite{Ge91} 
who introduces it as the method of multiple Markov chains. 
Independently it was developed in other papers. The {\it method of 
expanded ensembles} was introduced by Lyubartsev et al.~\cite{LyMa92} 
and proposes to enlarge the configuration space by introducing new 
dynamical variables. The better known {\it simulated 
tempering}~\cite{MaPa92} method of Marinari and Parisi
can be considered as the special
case where the temperature becomes the new dynamical variable. 
Building on simulated tempering the method of {\it parallel 
tempering} was introduced and tested by Hukusima and
Nemoto~\cite{HuNe96}. It is identical with Geyer's~\cite{Ge91}
proposal and well suited for clusters of workstations
and massively parallel computer architectures (although a 
moderately fast network between the nodes is sufficient).

Parallel tempering performs $n$ canonical MC simulations 
at different $\beta$-values with Boltzmann weight factors
$$ w_{B,i} (E^{(k)}) = e^{-\beta_i E^{(k)}},\ i=1, \dots , n,\
   \beta_1 < \beta_2 < ... < \beta_{n-1} < \beta_n $$
and allows exchange of neighbouring $\beta$-values
\begin{equation} \label{del_beta_PT}
 \beta_{i-1} \longleftrightarrow b_i\ ~~{\rm for}~~
 i=2, \dots , n\ .
\end{equation}
These transitions lead to the energy change
\begin{eqnarray} \nonumber
\triangle E = \left( -\beta_{i-1} E^{(k)}_i 
                    - \beta_i E^{(k')}_{i-1} \right)
 - \left( -\beta_i E^{(k)}_i 
 - \beta_{i-1} E^{(k')}_{i-1} \right) \\
 = \left( \beta_i-\beta_{i-1} \right)\, 
   \left( E^{(k)}_i - E^{(k')}_{i-1} \right)
\label{delE_PT}
\end{eqnarray}
which is accepted or rejected according to the Metropolis
algorithm. The $\beta_i$-values have to be determined
such that a reasonably large acceptance rate is obtained
for the $\beta$ exchange (\ref{del_beta_PT}) and 
ref.\cite{HuNe96} employs a recursive method due to  
Kerler and Rehberg~\cite{KeRe94}. This is similar to the
recursion needed at the beginning of a multicanonical simulation.

{\it Remark:} The method works for dynamical but not for statical
supercritical slowing down. Each member of the discrete set 
of weight factors samples still a Boltzmann distribution
({\it e.g.} it is not a valid method for calculating 
interfacial tensions).

\subsection{Random Walk Algorithms}

In ref.\cite{Be93} a class of algorithms was designed to perform a 
random walk in the energy (or any other function of the microstates). 
Assume the
Metropolis proposal probabilities $p^0(k',k)$ (\ref{p0}) are defined 
which allow for $N^{(k)}$ moves out of configuration $k$ (for the 
generalized Potts model probabilities (\ref{Potts_p0}) we have 
$N^{(k)}=(q-1) N$ independently of $k$). We may divide the $N^{(k)}$
moves into three classes: 
\begin{equation}
 N^{(k)} = N^+ + N^0 + N^-
\end{equation}
with

\begin{enumerate}

\item $N^+$ moves with $\triangle E_i > 0$, $i=1,\dots ,N^+$.

\item $N^0$ moves with $\triangle E_i = 0$, 
      $i=N^+ + 1,\dots ,N^+ + N^0$. 

\item $N^-$ moves with $\triangle E_i < 0$, 
      $i=N^+ + N^0 + 1,\dots ,N^{(k)} $. 

\end{enumerate}

The observation of~\cite{Be93} is that it is easy to define transition
probabilities $p^+_i$, $p^0_i$ and $p^-_i$,
$$ \sum_{i=1}^{N^+} p^+_i + \sum_{i=N^+ +1}^{N^+ +N^0} p^0_i +
   \sum_{i=N^+ + N^0 + 1}^{N^{(k)}} p^-_k  = 1\ , $$
such that
\begin{equation} \label{RWA}
  \sum_{i=1}^{N^+} p^+_i \triangle E^+_i =
- \sum_{i=N^+ +N^0 + 1}^{N^{(k)}} p^-_i \triangle E^-_i
\end{equation}
holds (but at the extrema). For any such choice the algorithm
performs (away from the extrema) a random walk in the energy and,
hence, samples a broad energy distribution. 

In \cite{Be93} configuration dependent transition probabilities 
\begin{equation} \label{pk}
 p_i^+ (k),\  p^0_i (k)\ ~~{\rm and}~~\  p^+_i (k)
\end{equation}
were used, where the dependence on the configurations $k$ exceeds
a mere energy dependence.
The algorithm sacrifices then an exact relationship with 
the canonical ensemble in favor of having a-priori well-defined 
defined transition probabilities and was
conjectured to be of interest for optimization problems, where one
focuses on minima and not in the canonical ensemble. With this
in mind the name {\it random cost algorithm} was chosen in~\cite{Be93}.

% A special case of equation~(\ref{RWA}) was later used in an attempt
% to calculate canonical expectation values~\cite{}. As pointed out
% in ref.\cite{} this is incorrect. Ref.\cite{} creates additional
% confusion with respect to this point although the author understood
% by then the argument quite well.

Wang's~\cite{Wa99} recent random walk algorithm is a special case of
equations~(\ref{RWA}), but uses microcanonical averages
\begin{equation} \label{pE}
 p_i^+ (E), \ p^0_i(E)\ ~~{\rm and}~~\ p^+_i(E)\ .
\end{equation}
The configuration dependence (\ref{pk}) is then reduced to energy 
dependence and canonical expectation values can be recovered by 
using the equation 
\begin{equation} \label{sp_eqn}
 n(E)\, N(E,\triangle E) = n(E+\triangle E)\, 
        N(E+\triangle E, -\triangle E)
\end{equation}
where $N(E,\triangle E)$ is the microcanonical average for the
number of transitions from configurations with energy $E$ to
configurations with energy $E'=E+\triangle E$. For proofs 
of~(\ref{sp_eqn}) see the appendix of~\cite{BeHa98} or 
ref.\cite{Ol98}.

It seems~\cite{Wa99} that in the random walk MC the use of estimators 
for $N(E,\triangle E)$ instead of their unknown exact values faces
more serious problems than in multicanonical simulations. Therefore, 
the best application of equation (\ref{sp_eqn})  might be to employ
it as input in the iteration towards the multicanonical weights
(\ref{wmuca}). Such a use of transition probabilities was first 
proposed by Smith and Bruce~\cite{SmBr95,SmBr96} 
and apparently deserves renewed attention.

%             Some interesting topics,
% like the dynamical-parameter method in $U(1)$ gauge theory~\cite{KeReWe95},
% had to be omitted because of space limitations. For the calculation
% of constraint effective potentials see Neuhaus in these proceedings.

\section{First Order Phase Transitions}

Multicanonical
simulations are best established for investigations of first-order 
phase transitions. The range of applications goes from studies of
mathematically ambitious topics to chemistry oriented ones. For
instance an investigation~\cite{BiNe93} of the mathematically
rigorous Borgs-Koteck\'y~\cite{BoKo91} FSS theory reveals that 
very strong phase transitions or very large lattices are needed 
to observe the asymptotic behavior predicted by their theory.
To give two examples from the chemistry side, an investigation of 
the coexistence curve of the Lennard-Jones fluid was performed in 
ref.\cite{Wi95,HuRe95} and liquid-vapor transitions in 
fluids were studied in ref.\cite{WiMu95,WiSc98}. The Lennard-Jones
fluid has also been a major field of applications for the 
original umbrella sampling method and variants 
thereof~\cite{ToVa77,DiVa93,Va93}.

A lot of work has focused on calculations of interfacial tensions and 
an over\-view is given in the forthcoming.

\subsection{$2d$ Potts Models}

The pioneering study was performed for the $2d$ 10-state Potts
model~\cite{BeNe92} and $2 f^s = 0.0978 (8)$ was found through FSS 
study of the equation (\ref{Itension}). Only afterwards the exact 
value was discovered to be $2f^s = 0.094701...$~\cite{BoJa92}. Once,
the exact result was known, the remaining, small discrepancy could be
eliminated by improving the finite volume estimators~\cite{BiNe94}. 
For these simulations the slowing down is around $\sim V^{2.3}$, 
{\it i.e.} reasonably close to the optimal performance.
For a related investigation of the 7-state $2d$ Potts model
see~\cite{JaBe92}.

\subsection{$2d$ and $3d$ Ising Model}

Many real physical systems fall into the universality class of the $3d$
Ising model. Despite its simplicity, it is therefore a very rewarding 
object to study.  Although many of its universal parameters have already
been determined with high precision, others are still 
in the making. In particular, there has been interest in the
universal surface tension and the critical-isotherm amplitude 
ratios~\cite{ZiFi96}. To obtain them, one needs accurate interfacial
tension results below the Curie temperature and multimagnetical 
simulations~\cite{BeHa93a,BeHa93b} have become the enabling technique
for Binder's~\cite{Bi82} histogram method, which was originally proposed
in this context.

For the $2d$ Ising model Onsager's exact result was reproduced with
good accuracy~\cite{BeHa93a}. However, for the $3d$ model the 
temperature dependence of the interfacial tension~\cite{BeHa93b} has, 
unfortunately, come out erratic. Therefore, 
the results of ref.\cite{HaPi94} appear to be the up-to-date best
estimates. Considerable technical improvements of multimagnetical
calculations are nowadays feasible and it seems worthwhile to start off 
a new multimagnetical $3d$ Ising model simulation.

\subsection{$SU(3)$ Gauge Theory}

One is interested in the interfacial tension for the
confinement/deconfinement phase transition. The use of multicanonical
techniques
has been explored by Grossmann, Laursen et al. \cite{GrLa92,GrLa93}.
In particular, they noticed that it is suitable to use an asymmetric
lattice, $V=L_z L^2 L_t$ with $L_z\ge 3L$. This forces the interfaces
into the $L^2$ plane and ensures a flat region for the minimum of 
equation  (\ref{Pmin}), thus greatly facilitating the extraction of 
finite-lattice values for the interfacial tension and, consequently,
the FSS analysis. 

For $SU(3)$ gauge theory the interfacial tension is
usually denoted by the symbol $\sigma$ and estimates are conveniently
given as multiples of $T_c^3=(T_c)^3$, where $T_c$ is the deconfinement 
temperature. Using the conventions of \cite{IwKa94}, the estimates
of~\cite{GrLa93} are
$$ \sigma = 0.052 (4)\, T_c^3,\ (L_t=2) ~~{\rm and}~~ 
   \sigma= 0.020 (2)\, T_c^3,\ (L_t=4) $$
This may be compared with the later estimate by Iwasaki et
al.~\cite{IwKa94}
$$ \sigma = 0.02925 (22)\, T_c^3,\ (L_t=4) ~~{\rm and}~~ 
   \sigma= 0.0218 (33)\, T_c^3,\ (L_t=6) $$
The discrepancy (only $L_t=4$ can be compared) is presumably due to
too small lattice sizes in~\cite{GrLa93}. Physically, one is interested
in the $L_t\to\infty$ limit. Possibly the strong $L_t$ dependence can 
be eliminated by using tadpole improved 
actions for which Beinlich et al. \cite{BeKa96} report
$$ \sigma = 0.0155 (16)\, T_c^3,\ (L_t=3 ~~{\rm\bf and}~~ L_t=4)\, .$$

\subsection{Electroweak Phase Transition}

Baryon violating processes are unsuppressed for $T>T_c$, where $T_c$
is the electroweak critical temperature. It has been conjectured, that
this may allow to explain the baryon asymmetry in nature. Models tie
the nucleation rate to the interface tension of the transition. Using
an effective scalar field theory~\cite{KaNePa95} or the full
theory~\cite{CsFo95}, multicanonical and related techniques turn out 
to be useful for simulations at a Higgs mass
$$ m_H \approx (35-37)\, GeV\, , $$
where one deals with a relatively strong first-order transitions, as
needed to explain the baryon asymmetry. Unfortunately, it turn out that
the transition weakens for higher Higgs masses, see ref.\cite{Mo95}
for a concise review.

\subsection{Weight factor estimates}

For spin systems with first-order phase transitions the 
FSS behavior is relatively well-known. Provided the steps
between system sizes are not too large, it is then possible 
to get working estimates of the $w_{mu}(E)$ weights by 
means of a FSS extrapolation from the already simulated smaller 
systems \cite{BeNe92,BeHa93b}. Another
method which works for these systems is patching of overlapping,
constraint \cite{Bh87} MC simulations. This has been employed by
various groups, but documentation has remained sketchy, presumably
it is best in~\cite{CsFo95}.
It should be emphasized that the purpose of the constrained
MC simulations is here to get a working estimate of the
multicanonical weights, whereas in~\cite{Bh87} they were used
for final estimates of physical observables. The advantage of
the multicanonical simulation is that it has not the ergodicity
problems from which microcanonical and constraint MC simulations
tend to suffer.  While these approaches work for first order phase
transitions, they fail as soon as ergodicity problems become severe
what is normally the case for complex systems. Then it is necessary 
to employ a recursion like the one of section~3, which is a
convenient choice for first order transitions too.

\section{Complex Systems}

Soon after the introduction of multicanonical methods, their potential
relevance for investigations of spin glasses and other complex systems
was pointed out~\cite{BeCe92}.
In these systems one encounters large free energy barriers due to
disorder and frustrations. Multicanonical simulations try to overcome 
the barriers through excursions into the disordered phase. Examples 
are spin glasses, proteins, hard optimization problems and others.

For an efficient application of the multicanonical idea one needs some 
some physical insight:  The configurations of interest have to be 
identified and a suitable parameterization has to be found which allows 
their actual enhancement. Whereas for first order phase transition
the appropriate parameters (temperature, magnetization, etc.) are
quite obvious, it is a major open problem whether suitable parameters
exist for classes of complex systems.

\subsection{Spin glasses}

Multicanonical studies have so far been limited to the simplest,
realistic prototypes, the $2d$ and $3d$ Edwards-Anderson $\pm J$ 
Ising\footnote{This is $J=0$ or $1$ in the 2-state Potts model
notation.}  spin glass~\cite{BeCe92,BeHa94}. 
Significant progress has been achieved with
respect to groundstate energy {\it and} entropy calculations. In
$3d$ the groundstate energy and entropy per spin ($e_g=E_g/N$ and
$s_g=S_g/N$) estimates of~\cite{BeHa94} are
$$ e_g = -1.8389 \pm 0.0040\ ~~~~{\rm and}~~\
   s_g = -1.7956 \pm 0.0042\ .$$
However, on the algorithmic side
the slowing down with volumes size is very bad, around $V^4$ or,
possibly, exponential. Certain advantages of simulated tempering are
claimed in~\cite{KeRe94} and~\cite{HuNe96}, but these studies fail
to present comparisons under identical conditions and presently
there is no firm evidence that simulated tempering yields a 
significantly better slowing down. These studies weight spin glass 
configurations with the inverse spectral density~(\ref{wspectral}). 
Recently a more focused application of the multicanonical method
to spin glass simulations was developed~\cite{BeJa98a}. The
idea is to explore the free-energy structure in the Parisi
order parameter such that the barriers in this variable can
be mapped out. These investigations are in progress. Marinari,
Parisi and collaborators perform major spin glass simulations
using parallel tempering, for a review see~\cite{MaPa97}.

\subsection{Proteins}

Proteins are linear polymers with the 20 naturally occurring amino 
acids as monomers. Chains smaller than a few tens of amino acids
are called peptides.
The problem is to predict the folded conformation of proteins and
peptides solely from their amino acid sequence. For many years the
emphasis of numerical investigations has been on finding the global
minimum potential energy and the major difficulty encountered is 
the multiple minima problem. Molecular dynamics has been the numerical
method of first choice, but the fraction of stochastic investigations 
shows an increasing trend. In particular simulated annealing has 
been frequently used. 

The major advantage of multicanonical and related methods in the 
context of proteins is that they allow for investigations of the 
{\it thermodynamics} of the entire free energy landscape of the
protein. This was realized by Hansmann and Okamoto~\cite{HaOk93}
when they introduced multicanonical sampling to the problem of
protein folding and, slightly later, by Hao and Scheraga~\cite{HaSc94}.
Since then numerous applications were performed and the 
simulations have been quite successful for peptides, but face
tremendous problems concerning scaling to large systems. A particularly
nice example is depicted in figure~5: The folding of poly-alanine
into its $\alpha$-helix coil~\cite{HaOk99a}. No \textit{a-priori}
information about the groundstate conformation is used in these
kind of simulations. By now a quite extensive literature exists
which is compiled in~\cite{HaOk99b}.

\begin{figure}[tb]
 \centerline{\hbox{ \psfig{figure=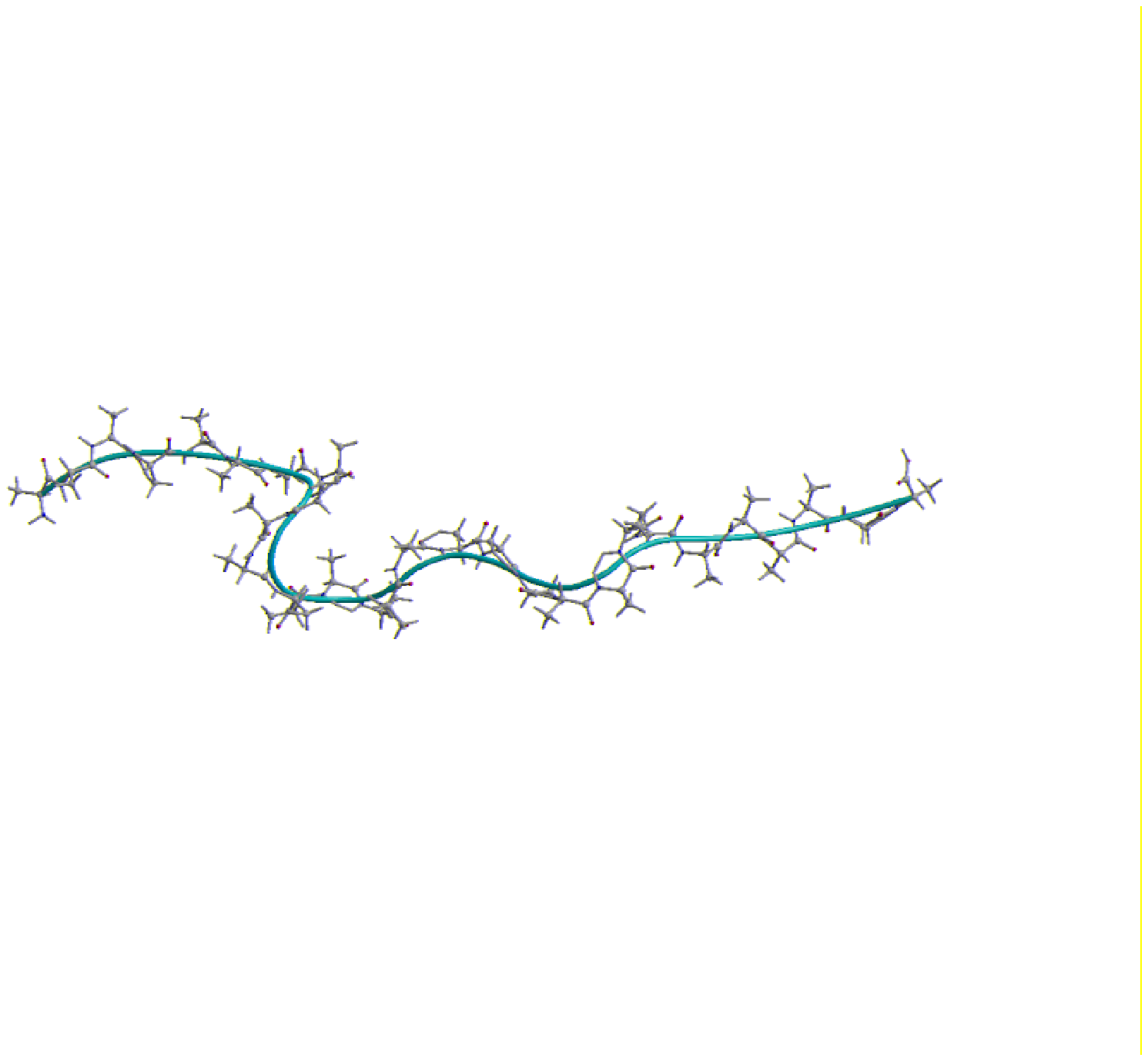,width=06cm} 
                    \psfig{figure=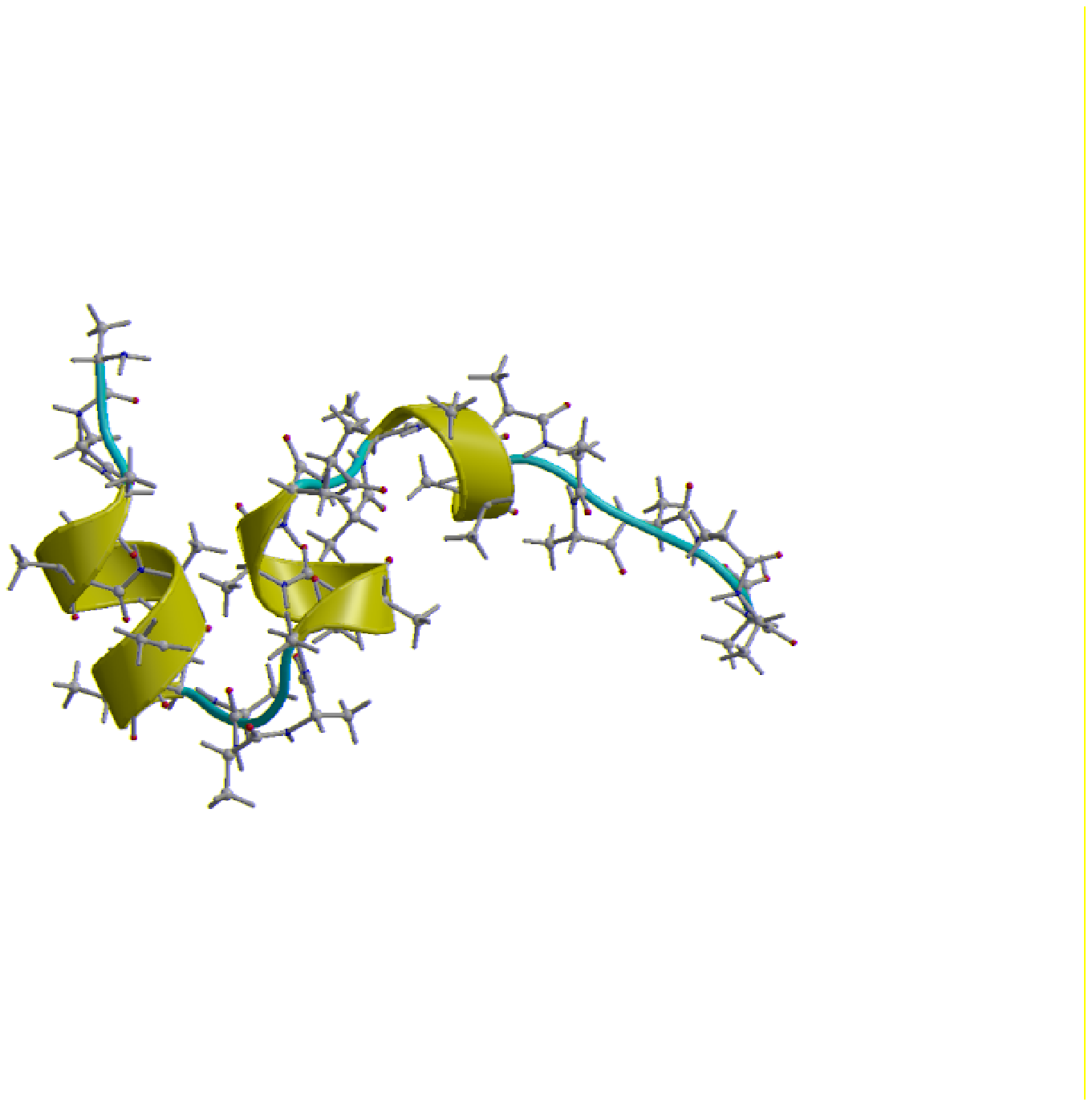,width=06cm} }}
 \centerline{\hbox{ \psfig{figure=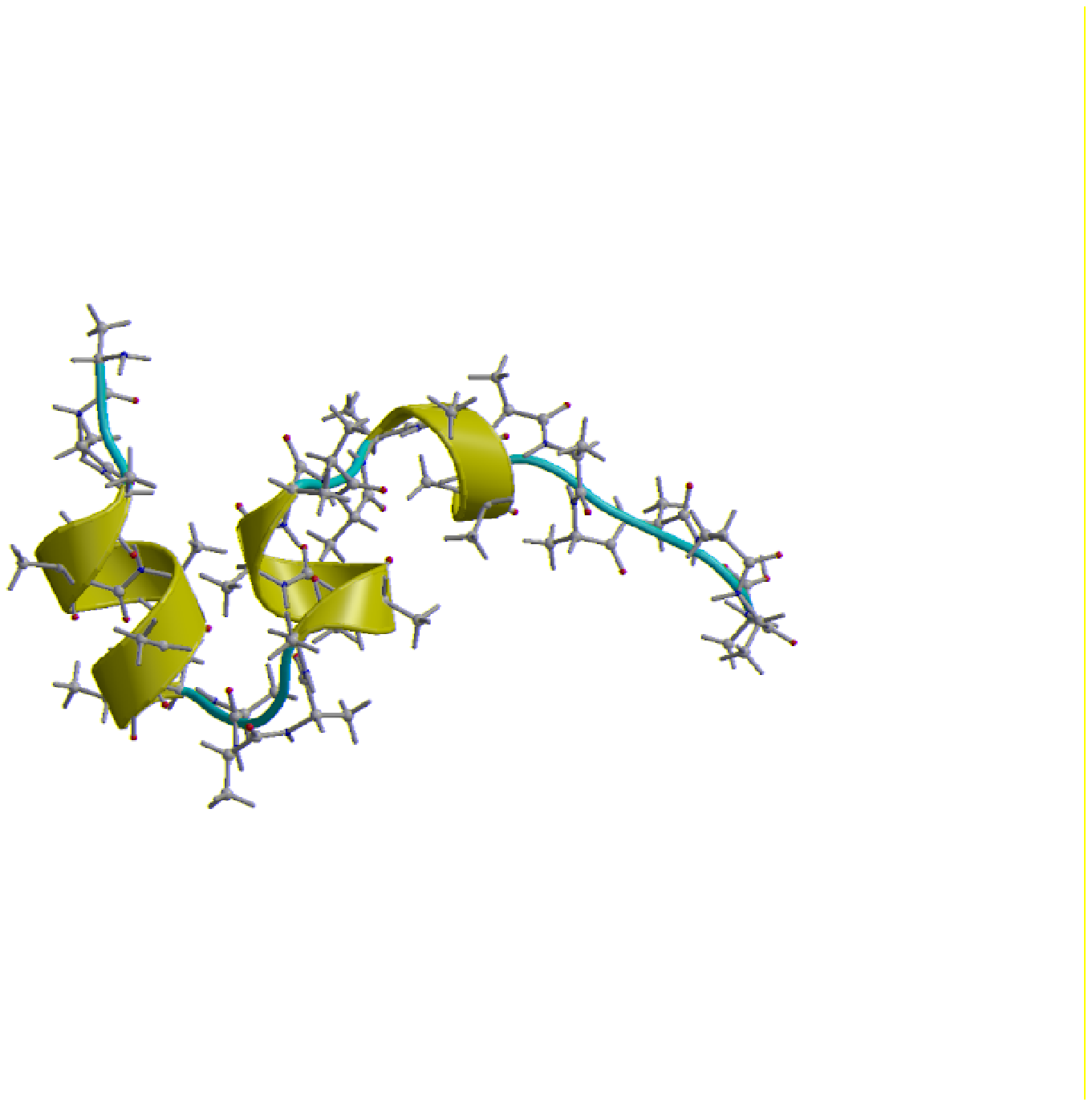,width=06cm} 
                    \psfig{figure=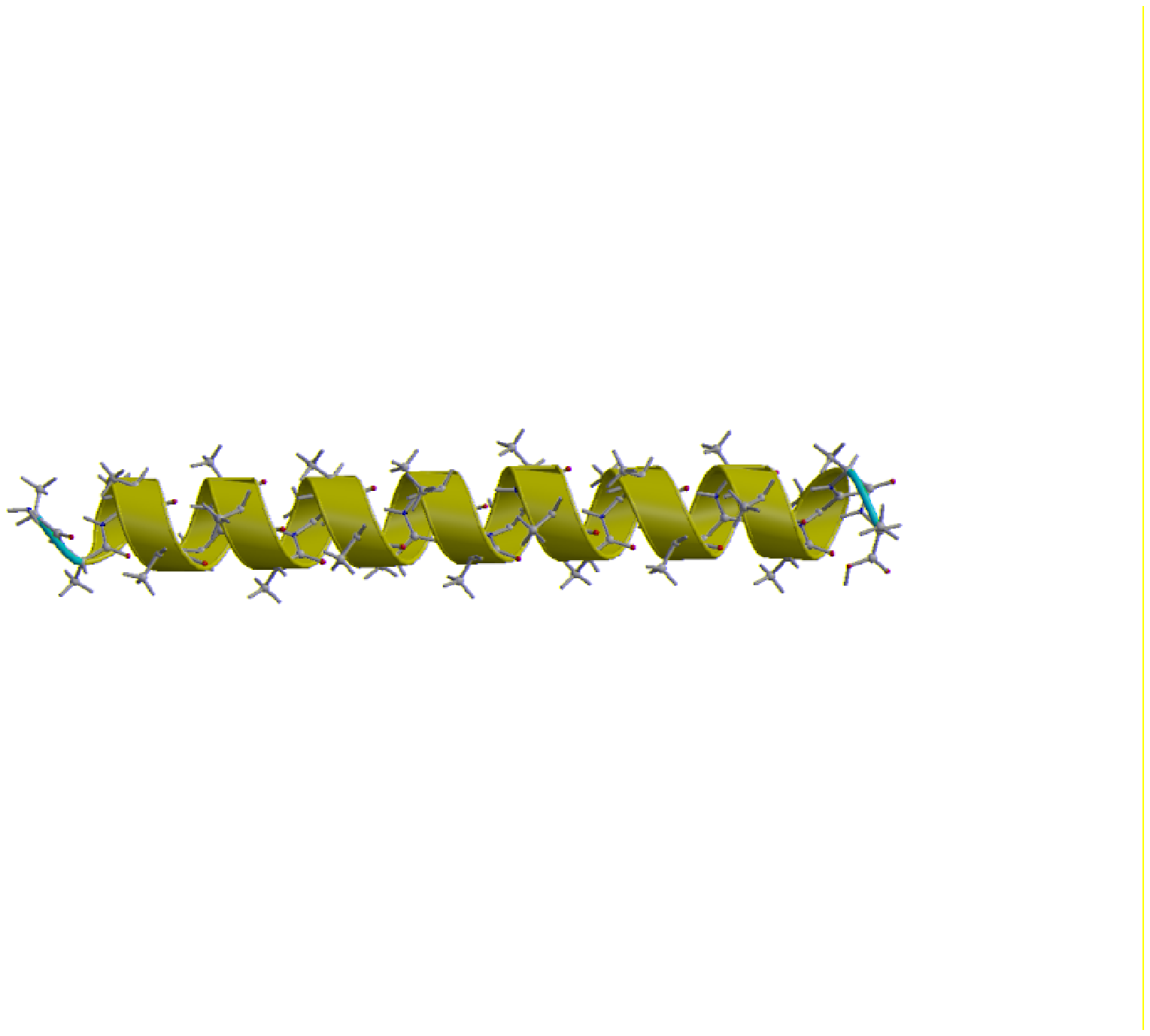,width=06cm} }}
 \caption{Example configurations from a multicanonical simulation
 of poly-alanine~\cite{HaOk99a} (courtesy Ulrich Hansmann).}
 \label{recur_bn}
\end{figure}

\subsection{Optimization problems}

They occur in engineering, network and chip design, traffic control, and
many other situations. General purpose algorithms for their solution are 
simulated annealing and genetic algorithms. To those, we may now like to 
add \textit{multicanonical annealing}~\cite{LeCh94} and 
\textit{random cost}~\cite{Be93}.

Multicanonical annealing is a combination of multicanonical 
sampling with variable
upper bounds and frequent adaption of parameters. A promising 
study~\cite{LeCh94} has been performed for the traveling salesman
problem. Up to $N=10,000$ cities, randomly distributed in the unit square,
were considered and scaling of the path length as function of $N$ was
investigated. The reported algorithmic performance is so good, that an 
independent confirmation would be desireable.

Recently the random cost algorithm (\ref{RWA}) has been applied to 
topology optimization in the engineering of trusses~\cite{BaKo99}.
The performance was found to be competitive to that of evolutionary
(genetic) algorithms.

\section{Outlook and Conclusions}

Sampling of broad energy distributions allows to overcome supercritical
slowing down. This is well established for first-order phase transitions.
Systems with conflicting constraints remain, despite some progress,
notoriously difficult and for them most hope lies on achieving further
algorithmic improvements. In addition, multicanonical methods may also 
be of interest when dealing with second order phase transitions, but so 
far nobody really cared to investigate this direction.

Multicanonical simulations have the potential to be replace canonical 
simulations as the method of first choice for exploratory MC studies. 
Their major advantage is that they give the thermodynamics for the 
entire temperature range of interest in one single MC run, whereas 
several to many (for each $\beta$-value one) canonical MC runs
are needed. So far the major stumbling block against using 
multicanonical simulations in this way has been the lack of easily 
available implementations of recursion schemes and data analysis 
programs. The author thinks that these shortcomings can be 
overcome.

\end{document}